# Negative energy density for a Dirac-Maxwell field

by

Dan Solomon

Rauland-Borg
3450 W. Oakton
Skokie, IL  60076
USA

Email: Dan.Solomon@Rauland.com

July 18, 1999

## Abstract

It is well know that there can be negative energy densities in quantum field theory. Most of the work done in this area has involved free non-interacting systems. In this paper we show how a quantum state with negative energy density can be formulated for a Dirac field interacting with an Electromagnetic field. It will be shown that for this case there exist quantum states whose average energy density over an arbitrary volume is a negative number with an arbitrarily large magnitude.

## I. Introduction.

Quantum field theory allows the existence of quantum states for which the energy density can be negative for a given point or region of space-time [1]. This differs from classical physics where the energy density is always positive. This result has important implications because the existence of negative energy density violates the weak energy condition of the General Theory of Relativity. This could, in principle, lead to the existence of various exotic phenomena, such as warp drives [2], wormholes, and closed time like curves [3], [4]. These phenomena are particularly disturbing because some of them (closed timelike curves, for example) can lead to causality violations.

Several authors have investigated the limits of the violation of the weak energy condition and have discovered that there may be certain bounds on the extent of this violation (see [5], [6], [7], [8], [9], [10], [11], [12]). However, these authors have only examined simple non-interacting quantum systems. Their results are not necessarily general and may apply only to the systems studied. Therefore, it is an open question on whether these bounds apply to more complicated quantum systems. In order to stimulate work in this area, it will be shown how to formulate quantum states with negative energy density for the Dirac-Maxwell field (i.e., a Dirac field interacting with an electromagnetic field). We will assume the Shrodinger representation so that the field operators are time independent and the state vectors obey the Schrodinger equation. Also throughout this discussion we assume natural units so that $\hbar = c = 1$.

## 2. Mathematical Details.

The energy density for a Dirac-Maxwell field is given by

$$\hat{T}^{00} = \frac{1}{2}\left(\left|\vec{\nabla}\times\hat{\vec{A}}\right|^2 + \left|\hat{\vec{E}}\right|^2\right) - \hat{\vec{J}}\cdot\hat{\vec{A}} + \hat{T}_D^{00} - T_R^{00} \tag{1}$$

where

$$\hat{T}_D^{00} = \hat{\psi}^\dagger\left(-i\vec{\alpha}\cdot\vec{\nabla} + m\beta\right)\hat{\psi} \tag{2}$$

(See Chapts. 5 and 6 of Greiner and Reinhardt [13]). In the above expressions $\hat{\vec{A}}$ is the vector potential operator, $\hat{\vec{E}}$ is the electric field operator, $\hat{\vec{J}}$ is the Dirac current operator, $\hat{\psi}$ is the Dirac field operator, and $T_R^{00}$ is a renormalization constant so that energy density of the vacuum state is zero. The energy density of a normalized state vector $|\Omega\rangle$ is given by $\langle\Omega|\hat{T}^{00}|\Omega\rangle$.

The energy density of the state $|\Omega\rangle$, averaged over a given volume V, is defined by

$$\xi_{ave}\left(|\Omega\rangle\right) = \frac{\int_V \langle\Omega|\hat{T}^{00}|\Omega\rangle d\vec{x}}{V} \tag{3}$$

where the integration is over the volume V.

We will quantize the electromagnetic field in the coulomb gauge (See Sakurai [14] and Ryder[15]). In this case

$$\vec{\nabla}\cdot\hat{\vec{A}} = 0 \tag{4}$$

and

$$\hat{\vec{E}} = \hat{\vec{E}}_\perp - \vec{\nabla}\hat{A}_0 \tag{5}$$

where $\hat{\vec{E}}_\perp$ is the operator for the transverse electric field and satisfies,

$$\vec{\nabla}\cdot\hat{\vec{E}}_{\perp} = 0 \tag{6}$$

$\hat{A}_0$ is the scalar potential operator which is given in terms of the Dirac charge operator $\hat{\rho}$ according to

$$\vec{\nabla}^2\hat{A}_0 = -\hat{\rho} \tag{7}$$

For the coulomb gauge the commutator relationships are given by

$$\left[\hat{A}^i(\vec{x}),\hat{\vec{E}}^j_{\perp}(\vec{x}')\right] = i\int\frac{d\vec{k}}{(2\pi)^3}\left(\delta^{ij} - \frac{k^i k^j}{|\vec{k}|^2}\right)e^{i\vec{k}\cdot(\vec{x}-\vec{x}')} \tag{8}$$

and

$$\left[\hat{E}^i_{\perp}(\vec{x}'),\hat{E}^j_{\perp}(\vec{x})\right] = \left[\hat{A}^i(\vec{x}'),\hat{A}^j(\vec{x})\right] = 0 \tag{9}$$

Also the Maxwell field operators $\hat{\vec{E}}_{\perp}(\vec{x})$ and $\hat{\vec{A}}(\vec{x})$ commute with the Dirac field operators $\hat{\psi}^{\dagger}$ and $\hat{\psi}$.

We will show that given a state vector $|\Omega\rangle$, whose current expectation value is non-zero at a given point $\vec{x}_0$, we can always formulate another state $|\Omega'\rangle$ whose energy density is negative at $\vec{x}_0$ where $|\Omega'\rangle$ is derived from $|\Omega\rangle$ according to the expression

$$|\Omega'\rangle = e^{i\hat{F}}|\Omega\rangle \tag{10}$$

where the operator $\hat{F}$ is defined by

$$\hat{F} = \int\hat{\vec{E}}_{\perp}\cdot\vec{\chi}\,d\vec{x} \tag{11}$$

and where $\vec{\chi}(\vec{x})$ is an real-valued vector function whose properties will be specified in the following discussion.

From the above expressions it shown in the Appendix that

$$\left[\hat{\vec{A}}(\vec{x}), \hat{F}\right] = i\left(\vec{\chi}(\vec{x}) + \frac{1}{4\pi}\int \frac{\vec{\nabla}\left(\vec{\nabla}\cdot\vec{\chi}(\vec{x}')\right)}{|\vec{x}-\vec{x}'|}d\vec{x}'\right) \tag{12}$$

Next we want to evaluate the commutator $\left[\hat{\vec{A}}(\vec{x}), e^{i\hat{F}}\right]$. Use the fact that $\left[\hat{\vec{A}}(\vec{x}), \hat{F}\right]$ commutes with $\hat{F}$ to obtain

$$\left[\hat{\vec{A}}(\vec{x}), \hat{F}^{n}\right] = \left[\hat{\vec{A}}(\vec{x}), \hat{F}\right] n\hat{F}^{n-1} \tag{13}$$

Use this in the following Taylor's expansion

$$e^{i\hat{F}} = 1 + i\hat{F} + \frac{i^2\hat{F}^2}{2!} + \ldots \frac{i^n\hat{F}^n}{n!} + \ldots \tag{14}$$

to obtain

$$\left[\hat{\vec{A}}(\vec{x}), e^{i\hat{F}}\right] = ie^{i\hat{F}}\left[\hat{\vec{A}}(\vec{x}), \hat{F}\right] \tag{15}$$

Note that $\hat{\vec{E}}_{\perp}$ is an observable and, therefore, hermitian. Therefore $\hat{\vec{E}}_{\perp}^{\dagger} = \hat{\vec{E}}_{\perp}$ so that $\hat{F}^{\dagger} = \hat{F}$. This yields,

$$e^{-i\hat{F}^{\dagger}} e^{i\hat{F}} = 1 \tag{16}$$

To simplify notation define

$$\vec{A}_{cl}(\vec{x}) = i\left[\hat{\vec{A}}(\vec{x}), \hat{F}\right] \tag{17}$$

Therefore

$$\vec{A}_{cl}(\vec{x}) = -\left(\vec{\chi}(\vec{x}) + \frac{1}{4\pi}\int \frac{\vec{\nabla}\left(\vec{\nabla}\cdot\vec{\chi}(\vec{x}')\right)}{|\vec{x}-\vec{x}'|}d\vec{x}'\right) \tag{18}$$

It is easy to show that

$$\vec{\nabla}\times\vec{A}_{cl}(\vec{x}) = -\vec{\nabla}\times\vec{\chi}(\vec{x}) \tag{19}$$

Use (15), (16) and (17) to show that

$$e^{-i\hat{F}^{\dagger}}\hat{\vec{A}}e^{i\hat{F}} = e^{-i\hat{F}^{\dagger}}\left(e^{i\hat{F}}\hat{\vec{A}} + \left[\hat{\vec{A}}, e^{i\hat{F}}\right]\right) = \hat{\vec{A}} + \vec{A}_{cl}(\vec{x}) \tag{20}$$

Also

$$\begin{aligned} e^{-i\hat{F}^{\dagger}}\hat{\vec{A}}(\vec{x})\hat{\vec{A}}(\vec{x}')e^{i\hat{F}} &= e^{-i\hat{F}^{\dagger}}\hat{\vec{A}}(\vec{x})e^{i\hat{F}}e^{-i\hat{F}^{\dagger}}\hat{\vec{A}}(\vec{x}')e^{i\hat{F}} \\ &= \left(\hat{\vec{A}}(\vec{x}) + \vec{A}_{cl}(\vec{x})\right)\left(\hat{\vec{A}}(\vec{x}') + \vec{A}_{cl}(\vec{x}')\right) \end{aligned} \tag{21}$$

Let $\hat{O} = \hat{\vec{J}}, \hat{E},$ or $\hat{T}_D^{00}$. Since $\hat{E}_{\perp}$, and therefore $\hat{F}$, commutes with all these quantities we have that

$$e^{-i\hat{F}^{\dagger}}\hat{O}e^{i\hat{F}} = \hat{O} \tag{22}$$

## 3. Negative Energy Density.

The energy density of the state $|\Omega'\rangle$ is

$$\langle\Omega'|\hat{T}^{00}|\Omega'\rangle = \langle\Omega|e^{-i\hat{F}^{\dagger}}\hat{T}^{00}e^{i\hat{F}}|\Omega\rangle \tag{23}$$

where we have assumed that $|\Omega\rangle$ is normalized which implies that $|\Omega'\rangle$ is normalized.

From the previous discussion this yields

$$\langle\Omega'|\hat{T}^{00}|\Omega'\rangle = \langle\Omega|\left\{\frac{1}{2}\left(\left|\vec{\nabla}\times\left(\hat{\vec{A}} + \vec{A}_{cl}\right)\right|^2 + \left|\hat{\vec{E}}\right|^2\right) - \hat{\vec{J}}\cdot\left(\hat{\vec{A}} + \vec{A}_{cl}\right) + \hat{T}_D^{00} - T_R^{00}\right\}|\Omega\rangle \tag{24}$$

Note, the effect of the operation $e^{-i\hat{F}^{\dagger}}\hat{T}^{00}e^{i\hat{F}}$ is to replace all $\hat{\vec{A}}$ with $\hat{\vec{A}} + \vec{A}_{cl}$.

Rearrange terms to obtain

$$\langle\Omega'|\hat{T}^{00}|\Omega'\rangle = \langle\Omega|\left\{\frac{1}{2}\left(\left|\vec{\nabla}\times\hat{\vec{A}}\right|^2 + \left|\hat{\vec{E}}\right|^2\right) - \hat{\vec{J}}\cdot\hat{\vec{A}} + \hat{T}_D^{00} - T_R^{00}\right\}|\Omega\rangle$$
$$+\langle\Omega|\left\{\frac{1}{2}\left|\vec{\nabla}\times\vec{A}_{cl}\right|^2 + \left(\vec{\nabla}\times\hat{\vec{A}}\right)\cdot\left(\vec{\nabla}\times\vec{A}_{cl}\right) - \hat{\vec{J}}\cdot\vec{A}_{cl}\right\}|\Omega\rangle \tag{25}$$

Use (1) in the above, to obtain

$$\langle\Omega'|\hat{T}^{00}|\Omega'\rangle = \hat{T}_e^{00} + \frac{1}{2}\left|\vec{\nabla}\times\vec{A}_{cl}\right|^2 + \left(\vec{\nabla}\times\vec{A}_e\right)\cdot\left(\vec{\nabla}\times\vec{A}_{cl}\right) - \vec{J}_e\cdot\vec{A}_{cl} \tag{26}$$

where $\vec{A}_e$, $\vec{J}_e$, and $\hat{T}_e^{00}$ are the expectation values of the vector potential, the current density, and the energy density, respectively, for the state vector $|\Omega\rangle$. For a normalized $|\Omega\rangle$ these quantities are defined by

$$\vec{A}_e = \langle\Omega|\hat{\vec{A}}|\Omega\rangle \tag{27}$$

$$\vec{J}_e = \langle\Omega|\hat{\vec{J}}|\Omega\rangle \tag{28}$$

and

$$\hat{T}_e^{00} = \langle\Omega|\hat{T}^{00}|\Omega\rangle \tag{29}$$

Use (19) in (26) to obtain

$$\langle\Omega'|\hat{T}^{00}|\Omega'\rangle = \hat{T}_e^{00} + \frac{1}{2}\left|\vec{\nabla}\times\vec{\chi}\right|^2 - \left(\vec{\nabla}\times\vec{A}_e\right)\cdot\left(\vec{\nabla}\times\vec{\chi}\right) - \vec{J}_e\cdot\vec{A}_{cl}(\vec{x};\vec{\chi}) \tag{30}$$

where we write $\vec{A}_{cl}(\vec{x};\vec{\chi})$ to denote the dependence of $\vec{A}_{cl}$ on the function $\vec{\chi}$. Note that the quantities $\hat{T}_e^{00}$, $\vec{A}_e$, and $\vec{J}_e$ are independent of the function $\vec{\chi}$. Therefore we can manipulate $\vec{\chi}$ without effecting these quantities.

Now, for a given $|\Omega\rangle$, how can we determine $\vec{\chi}$ so that $\langle\Omega'|\hat{T}^{00}|\Omega'\rangle$ is negative at a given point $\vec{x}_0$? First assume that $\vec{J}_e(\vec{x}_0) \neq 0$. Next, find a function $\vec{\chi}_1(\vec{x})$ so that at the point $\vec{x}_0$ the following conditions are satisfied

$$\vec{\nabla} \times \vec{\chi}_1(\vec{x}_0) = 0 \text{ and } \vec{J}_e(\vec{x}_0) \cdot \vec{A}_{cl}(\vec{x}_0; \vec{\chi}_1) \neq 0 \tag{31}$$

Let $\vec{\chi}(\vec{x}) = f\vec{\chi}_1(\vec{x})$ where f is a real valued constant . Also note that $\vec{A}_{cl}(\vec{x}; f\vec{\chi}_1) = f\vec{A}_{cl}(\vec{x}; \vec{\chi}_1)$. Use this in (30) to obtain,

$$\langle\Omega'|\hat{T}^{00}(\vec{x}_0)|\Omega'\rangle = \hat{T}_e^{00}(\vec{x}_0) - f\vec{J}_e(\vec{x}_0) \cdot \vec{A}_{cl}(\vec{x}_0; \vec{\chi}_1) \tag{32}$$

From the form of this equation it is obvious that it is always possible to find an f such that

$$\langle\Omega'|\hat{T}^{00}(\vec{x}_0)|\Omega'\rangle < 0 \tag{33}$$

Now let us consider the energy density averaged over a given volume V for the state $|\Omega'\rangle$. Find a function $\vec{\chi}_1(\vec{x})$ that satisfies the following conditions

$$\vec{\nabla} \times \vec{\chi}_1(\vec{x}_0) = 0 \text{ over the volume V} \tag{34}$$

and

$$\int_V \vec{J}_e(\vec{x}) \cdot \vec{A}_{cl}(\vec{x}; \vec{\chi}_1) d\vec{x} \neq 0 \tag{35}$$

The first of the above conditions can always be met by making $\vec{\chi}_1(\vec{x})$ constant over the volume V. Given the general nature of $\vec{\chi}_1(\vec{x})$ we will assume that we can always find a quantum state $|\Omega\rangle$ and function $\vec{\chi}_1(\vec{x})$ where (34) and (35) hold. Now, as before, let $\vec{\chi}(\vec{x}) = f\vec{\chi}_1(\vec{x})$. Then

$$\xi_{ave}(|\Omega'\rangle) = \frac{\int_V \langle\Omega'|\hat{T}^{00}|\Omega'\rangle d\vec{x}}{V} = \frac{\int_V \left(\hat{T}_e^{00} - f\vec{J}_e \cdot \vec{A}_{cl}(\vec{x}; \vec{\chi}_1)\right) d\vec{x}}{V} \tag{36}$$

This yields

$$\xi_{ave}\left(\left|\Omega'\right\rangle\right)=\xi_{ave}\left(\left|\Omega\right\rangle\right)-f\,\frac{\int_V\left(\vec{J}_e\cdot\vec{A}_{cl}\left(\vec{x};\vec{\chi}_1\right)\right)d\vec{x}}{V} \tag{37}$$

We can always find an f which makes $\xi_{ave}\left(\left|\Omega'\right\rangle\right)$ a negative number with an arbitrarily large magnitude.

## Conclusion

It has been shown how to formulate a quantum state with negative energy density for a Dirac-Maxwell field. It has been shown that the average energy density over some volume V can be negative with an arbitrarily large magnitude. There are no restrictions on the size of the volume. Therefore, it is possible to formulate quantum states, whose total energy within an arbitrarily large volume, is a negative number with an arbitrarily large magnitude.

## Appendix

From the definition of $\hat{F}$

$$\left[\hat{\vec{A}}(\vec{x}),\hat{F}\right]=\int\left[\hat{\vec{A}}(\vec{x}),\hat{\vec{E}}_\perp(\vec{x}')\right]\cdot\vec{\chi}(\vec{x}')d\vec{x}' \tag{A1}$$

Use (8) in the above to obtain

$$\left[\hat{\vec{A}}(\vec{x}),\hat{F}\right]=i\int d\vec{x}'\int\frac{d\vec{k}}{(2\pi)^3}\left(\vec{\chi}(\vec{x}')-\frac{\vec{k}\left(\vec{k}\cdot\vec{\chi}(\vec{x}')\right)}{\left|\vec{k}\right|^2}\right)e^{i\vec{k}\cdot(\vec{x}-\vec{x}')} \tag{A2}$$

Use,

$$\int\frac{d\vec{k}}{(2\pi)^3}e^{i\vec{k}\cdot(\vec{x}-\vec{x}')}=\delta^{(3)}(\vec{x}-\vec{x}')$$

to obtain

$$\left[\hat{\vec{A}}(\vec{x}),\hat{F}\right]=i\vec{\chi}(\vec{x})-i\int d\vec{x}'\int\frac{d\vec{k}}{(2\pi)^3}\frac{\vec{k}\left(\vec{k}\cdot\vec{\chi}(\vec{x}')\right)}{\left|\vec{k}\right|^2}e^{i\vec{k}\cdot(\vec{x}-\vec{x}')} \tag{A3}$$

Operate on the above with $\vec{\nabla}^2$ to obtain

$$\vec{\nabla}^2\left[\hat{\vec{A}}(\vec{x}),\hat{F}\right]=i\vec{\nabla}^2\vec{\chi}(\vec{x})+i\int d\vec{x}'\int\frac{d\vec{k}}{(2\pi)^3}\vec{k}\left(\vec{k}\cdot\vec{\chi}(\vec{x}')\right)e^{i\vec{k}\cdot(\vec{x}-\vec{x}')} \tag{A4}$$

This yields

$$\vec{\nabla}^2\left[\hat{\vec{A}}(\vec{x}),\hat{F}\right]=i\vec{\nabla}^2\vec{\chi}(\vec{x})-i\int d\vec{x}'\int\frac{d\vec{k}}{(2\pi)^3}\vec{\nabla}_{\vec{x}}\left(\vec{\chi}(\vec{x}')\cdot\vec{\nabla}_{\vec{x}}\right)e^{i\vec{k}\cdot(\vec{x}-\vec{x}')} \tag{A5}$$

From the above we obtain

$$\vec{\nabla}^2\left[\hat{\vec{A}}(\vec{x}),\hat{F}\right]=i\left(\vec{\nabla}^2\vec{\chi}(\vec{x})-\vec{\nabla}\left(\vec{\nabla}\cdot\vec{\chi}\right)\right) \tag{A6}$$

Use the fact that the solution of the expression

$$\vec{\nabla}^2 G=-\sigma \tag{A7}$$

is

$$G=\int\frac{\sigma(\vec{x}')}{4\pi|\vec{x}'-\vec{x}|}d\vec{x}' \tag{A8}$$

in (A6) to obtain

$$\left[\hat{\vec{A}}(\vec{x}),\hat{F}\right]=i\left(\vec{\chi}(\vec{x})+\int\frac{\vec{\nabla}\left(\vec{\nabla}\cdot\vec{\chi}(\vec{x}')\right)}{4\pi|\vec{x}'-\vec{x}|}d\vec{x}'\right) \tag{A9}$$